\documentclass[journal]{IEEEtran}
\IEEEoverridecommandlockouts 

\makeatletter
\let\NAT@parse\undefined
\makeatother

\usepackage{amsmath,amssymb,amsfonts,amsthm,commath,esint}
\usepackage{bm} 
\usepackage[numbers,sort&compress]{natbib}

\usepackage{mathtools}
\usepackage{multirow}
\usepackage{physics}
\usepackage{color} 

\usepackage{graphicx}  
\usepackage{enumitem}  
\usepackage{optidef} 
\usepackage{mleftright}
\usepackage{cuted}
\usepackage{algorithm2e}
\usepackage{microtype}
\usepackage{ifthen} 
\usepackage{subcaption}
\usepackage{soul}

\SetKwInput{kwInit}{Initialization}
\ifCLASSOPTIONcompsoc
    \usepackage[caption=false, font=normalsize, labelfont=sf, textfont=sf]{subfig}
\else
\usepackage[caption=false, font=footnotesize]{subfig}
\fi

\makeatletter
\def\blfootnote{\xdef\@thefnmark{}\@footnotetext}
\makeatother

\newcommand{\iu}{\mathrm{i}\mkern1mu} 

\newcommand{\es}[1]{\sin{\left(#1\right)}}
\newcommand{\ec}[1]{\cos{\left(#1\right)}}
\newcommand{\p}[1]{\left(#1\right)}

\renewcommand{\vec}[1]{\mathbf{\lowercase{#1}}}

\newcommand{\veci}[1]{\pmb{\lowercase{#1}}}

\newcommand{\T}{^\mathsf{T}}    

\newcommand{\e}[1]{\mathrm{e}^{#1}}

\DeclarePairedDelimiterXPP\Aver[1]{\mathbb{E}}{[}{]}{}{

#1
}

\begin{document}

\title{Prediction of Wireless Channel Statistics with \\ Ray Tracing and Uncalibrated Digital Twin\\ 
\thanks{Mahmoud Abouamer (mabouamer@uwaterloo.ca; mahmoudabo@es.aau.dk), Robin J. Williams (rjw@es.aau.dk) and Petar Popovski (petarp@es.aau.dk) are with the Department of Electronic Systems, Aalborg University, Denmark. Mahmoud Abouamer and Robin J. Williams contributed equally to this work. Mahmoud Abouamer was with University of Waterloo, Canada. \\
This research is supported by the HORIZON JU-SNS-2022-STREAM-B-01-02 CENTRIC project (Agrmt. No. 101096379) and the Velux Foundation, Denmark, through the Villum Investigator Grant WATER (Agrmt. No. 37793).
}

}
\author{\IEEEauthorblockN{Mahmoud Saad Abouamer,~\IEEEmembership{Graduate Student Member,~IEEE}, Robin J. Williams, Petar Popovski,~\IEEEmembership{Fellow,~IEEE}}}

\maketitle

\begin{abstract}
We introduce a framework for predicting wireless channel statistics based on digital twin (DT) and ray tracing. The DT is derived from satellite images and is \emph{uncalibrated}, as it does not assume precise information on the electromagnetic properties of the materials in the environment. The uncalibrated DT is utilized to derive a geometric prior that informs a Gaussian process (GP) and thereby predict channel statistics using only a few measurements. The framework also quantifies uncertainty, offering statistical guarantees for rate selection in ultra-reliable low-latency communication (URLLC). Experimental validation demonstrates the efficacy of the proposed framework using measurement data. 
\end{abstract}


\section{Introduction}\label{sec:introduction}
Digital twins (DTs) that accurately model wireless channels have demonstrated significant utility for both communication and localization tasks \cite{survey_DT}. For instance, by maintaining a precise DT, site-specific digital twins can be used to reduce CSI acquisition and feedback overhead \cite{AK_DT_communication}. However, for the DT to provide accurate predictions,  its electromagnetic properties should match real-world channel conditions. 

The process of configuring the objects and materials in the DT to represent the wireless environment is referred to as \emph{calibration}. Differentiable ray tracers (RT), such as Sionna RT \cite{Sionna_RT} are instrumental in enabling effective calibration methods.  Differentiable RTs allow the computation of gradients for system parameters, enabling tasks like data-driven material property optimization \cite{RT_calibration}.

This work addresses resource allocation for ultra-reliable low-latency communication (URLLC) communications, aiming to select a data rate that meets a statistical reliability constraint. Due to latency limitations, channel feedback may be infeasible and hence the system must rely on alternatives such as channel statistics maps to select a rate which satisfies the reliability constraint with a sufficiently high probability. While a fully calibrated DT can directly predict channel statistics by accurately modeling objects and materials, frequent calibration is resource-intensive and limits its practical use. To address this, we focus on exploiting uncalibrated DTs that can be directly derived from satellite images available through open-source maps like OpenStreetMap \cite{OpenStreetmaps}, without relying on prior knowledge of material properties in the scene or requiring data-intensive calibration processes.

We summarize our contributions and results as follows. An uncalibrated DT is used to extract geometric features of the environment. These are combined with channel measurements from a few anchor points to estimate the channel statistics across the entire scene. The estimation leverages the spatial interpolation capabilities of a Gaussian process (GP) along with the geometric consistency of the DT to predict channel statistics with high accuracy in the region of interest. This is validated via experimental data obtained through a measurement campaign. The results show that the proposed geometry-informed GP significantly improves prediction accuracy across the entire scene compared to the benchmarks where \emph{(i)} an uncalibrated DT directly predicts channel statistics; and \emph{(ii)} a GP uses location information to predict channel statistics. Moreover, when the channel statistics map is used in rate selection for URLLC, the proposed framework is shown to,  approximately, double the  data rate while maintaining the same reliability, compared to the benchmarks. In summary, the geometry-informed GP framework scales the small number of calibration measurements to channel predictions that are statistically accurate for the entire region of interest.

 \subsubsection*{Organization} Section~\ref{sec:system_model} presents the problem formulation, and the construction of the uncalibrated DT.  Section~\ref{sec: Site-specific Statistics Prediction} presents the channel-prediction framework, and Section~\ref{sec:scenario} presents an effective method for experimentally measuring channel-fading statistics. Finally, Section~\ref{sec:simulation} details the rate-selection framework and its experimental validation.

\section{System model and Problem Formulation} \label{sec:system_model}
We consider a system where an access point (AP) serves a user (UE) located at position $\vec{x}$. The UE, equipped with a single antenna, acts as the transmitter (TX) and transmits its zero-mean, unit-power symbol $s \in \mathbb{C}$  with transmit power $P_\text{tx} \geq 0$. The single-antenna AP then functions as the receiver (RX). The received narrowband signal, after delay-spread equalization, is 
\begin{align}
  y = \sqrt{P_\text{tx}} h\p{\vec{x}} s + n,  
\end{align}
where $h\p{\vec{x}} \in \mathbb{C}$ is the channel between the transmitter and receiver, and $n \sim \mathcal{CN}\p{0, \sigma_n^2}$ is additive white Gaussian noise (AWGN). Thus, the SNR given by $\gamma\p{\vec{x}} =  \frac{P_\text{tx} \abs{h\p{\vec{x}}}^2}{\sigma_n^2}$.

\vspace{0.1cm}

\subsection{Problem formulation}
{For a user with position $\mathbf{x}$, the goal is to select the maximum rate $R(\vec{x})$ such that the outage probability is at most $\epsilon \in (0,1)$. {The outage probability represents the probability of exceeding the channel capacity, $p_{\text{out}, \vec{x}} = P \big({\log_2({1 + \gamma({\vec{x}}})) < R(\vec{x})}\big)$.} If  the distribution of $h\p{\vec{x}}$ is known, then $R(\vec{x})$ can be selected using the statistics of the channel power $p(\vec{x})= |h\p{\vec{x}}|^2$. However, due to latency constraints and the dynamic properties of the wireless communication channel, the distribution of $p(\vec{x})$ is not perfectly known. To account for the uncertainty associated with $p_{\text{out}, \vec{x}}$}, this work aims to maximize the rate $R (\vec{x})$ that can be selected while ensuring compliance with a reliability constraint that bounds the meta-probability as $\tilde{p}_{\epsilon} \overset{\Delta}{=}P\p{ p_\text{out,$\vec{x}$} > \epsilon  } \leq \delta$. This can also be expressed as \cite{kallehauge2024experimental}
\begin{align}
\label{meta_probability}
 \tilde{p} _{\epsilon}= P\p{ R(\vec{x}) > \log_2\p{1 +  \frac{P_\text{tx}  p_\epsilon\p{\vec{x}}}{\sigma_n^2} } } \leq \delta.  
\end{align}
Here, $p_\epsilon\p{\vec{x}}$ is the $\epsilon$-quantile the channel power. In this work, we rely on statistical radio maps to estimate the power quantiles $p_\epsilon\p{\vec{x}}$. To this end, we employ an uncalibrated digital twin (DT) as a geometric prior to inform a Gaussian process (GP), enabling channel-statistic predictions from only a few measurements.

Traditional radio maps require dense measurements and lack geometric context. On the other hand, while capturing the underlying geometry, uncalibrated digital twins yield inaccurate channel statistics, as ray tracing relies on assumed electromagnetic properties that may not reflect real materials. This mismatch distorts the received power, impacting the prediction of URLLC-relevant statistics (i.e., lower quantiles). To address these limitations, our proposed framework, presented in Section~\ref{sec: Site-specific Statistics Prediction}, leverages geometry-based DT outputs to embed spatial correlations within a GP, which utilizes a small set of measurements to accurately predict the channel statistics. GPs are employed here for their sample-efficient spatial interpolation properties, enabling not only efficient prediction but also closed-form uncertainty quantification \cite{Gp_book}.

Subsequently, in contrast to conventional radio maps that provide point estimates of average statistics, the proposed framework provides predictions of URLLC-relevant statistics along with closed-form quantification of prediction uncertainty. This enables the selection of transmission rates under statistical guarantees. In Section~\ref{sec:simulation}, we validate our GP-based framework using real measurements, demonstrating its ability to accurately predict channel statistics and meet URLLC reliability targets.

\begin{figure}[t]
    \centering
    \includegraphics[width=\linewidth, height= 5.5cm]{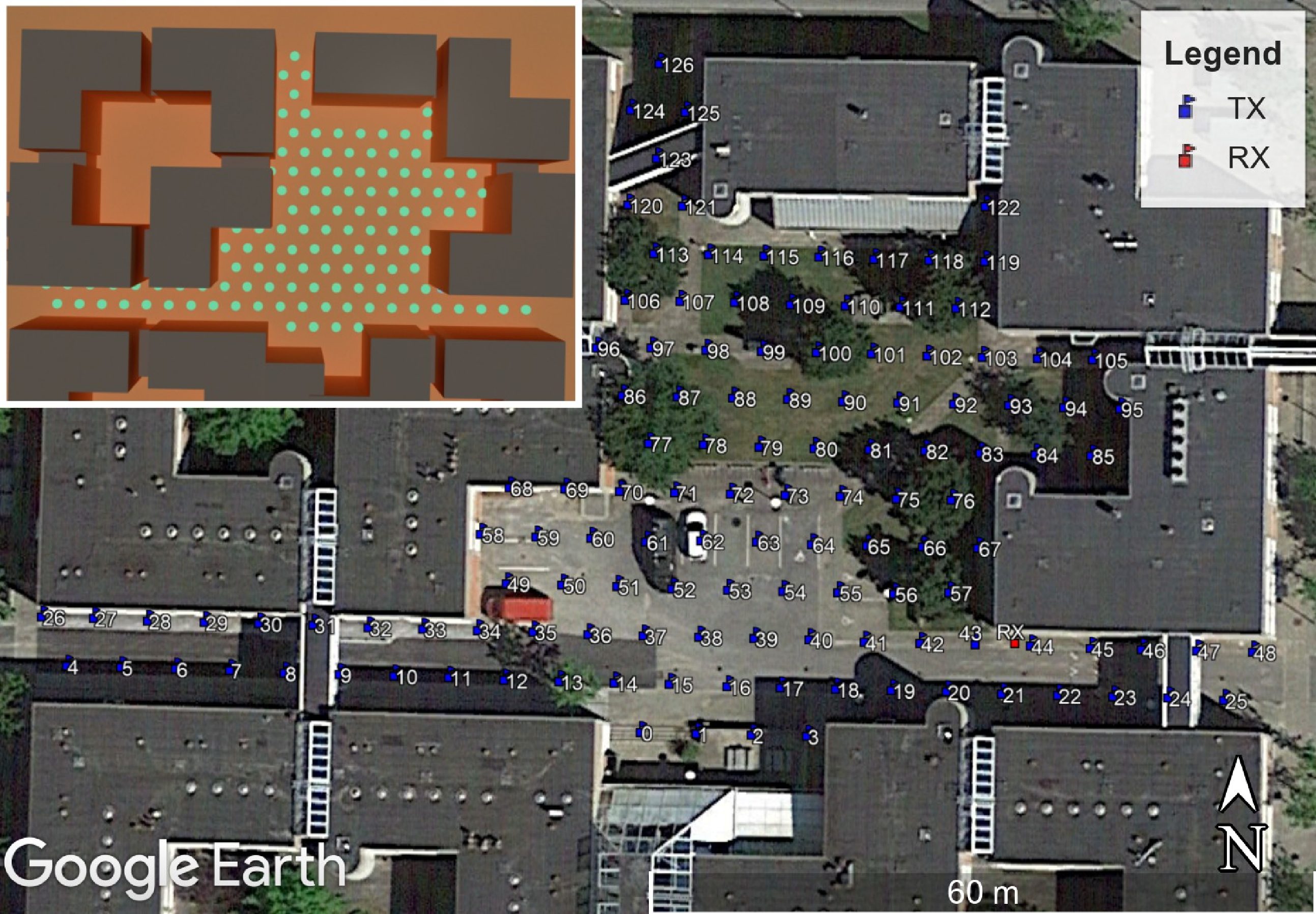}
    \caption{\textit{Bottom right}: A satellite picture of the outdoor scenario not taken at the day of the measurement. Transmitter locations marked with blue dots \cite{kallehauge2024experimental}. \textit{Top left}: Illustration of the geometry made using the \texttt{sionna.rt.scene.render} python library \cite{Sionna_RT} and data extracted from OpenStreetMap \cite{OpenStreetmaps}. Transmitter locations are marked as green dots.}
    \label{fig:scene_layout}
\end{figure}

\subsection{Scene Description and Digital‐Twin Formation}
\label{section: scene description}
We consider the scenario illustrated in Fig.~\ref{fig:scene_layout} (Left), featuring a stationary receiver and 127 distinct transmitter locations, each equipped with an omnidirectional antenna. A digital twin (DT) of this environment was created by extracting the geometry from OpenStreetMap\cite{OpenStreetmaps} using the \emph{Blosm for Blender} plugin, and importing it into the Sionna Python library\cite{Sionna_RT}. In the DT (Fig.~\ref{fig:scene_layout} (Right)), the 127 transmitter sites are marked as green dots. All surfaces, including the ground, are modeled with Sionna’s predefined \texttt{itu\_brick} material (relative permittivity $\epsilon_r = 3.91$) \cite[Table~3]{ITU-R_P.2040-3}. No calibration of the DT is performed.

\section{Site-specific Channel Statistics Prediction}  \label{sec: Site-specific Statistics Prediction}
In this section, we focus on channel statistics prediction, with the rate-selection problem addressed and validated using experimental data in Section~\ref{section: Rate selection under reliability constraints}. 
 
\subsection{Data Preprocessing and Estimation of Quantiles}
Consider an input dataset $\mathcal{\hat{D}}=\{\vec{p}_d,\vec{x}_d\}_{d=1}^D$, where for each location $\vec{x}_d$, $\vec{p}_d=  \left( p_{d,1}, \ \ldots ,\ p_{d,N} \right)$ captures $N$ independent small-scale fading channel measurements. To enable spatial channel prediction, the logarithm of the $\epsilon$-quantile channel power, defined as $q_\epsilon\p{\vec{x}} = \ln\p{p_\epsilon\p{\vec{x}}}$, is modeled as a  Gaussian process (GP). The logarithmic transformation is applied to allow transformed data to be effectively modeled as a GP. To estimate the log $\epsilon$-quantile $q_{\epsilon}(\vec{x})$,
the log-fading power quantile at a location $ \vec{x}_d$ is estimated as $\widehat{q}_{\epsilon,d} = \ln\left({p}_{d,(r)}\right)$ for $r = \lfloor N\epsilon \rfloor$ and $p_{d,(r)}$ denoting the $r$-th order statistic of $\vec{p}_d$. {Indeed, this estimate is
unbiased and asymptotically behaves as a Gaussian distribution \cite{kallehauge2022predictive}.} Subsequently, the input to the  channel prediction scheme is a dataset $\mathcal{{D}}=\{\widehat{q}_{\epsilon,d},\vec{x}_d\}_{d=1}^D$.

\subsection{GP for Channel Statistics Prediction}

By modeling the quantile function $q_\epsilon\p{\vec{x}} $ as a Gaussian process, GP regression can be used to predict fading statistics by  interpolating channel measurements resulting in a predictive distribution.  {This approach was proposed in \cite{kallehauge2022predictive} to learn a predictive distribution $q_{\epsilon}^{\sf{spatial}} (\vec{x}| \mathcal{D}) \sim \mathcal{N} (\mu(\vec{x}| \mathcal{D}), \sigma^{2} (\vec{x}| \mathcal{D})),$ where $\mu(\vec{x}| \mathcal{D})$ is the {predictive mean} and $\sigma^2(\vec{x}| \mathcal{D})$ is the predictive variance.} For the implementation details, we refer the reader to \cite[Sec. III-B]{kallehauge2022predictive}. Notably, in this work, we go beyond spatial interpolation by incorporating geometric correlations extracted from an uncalibrated DT to enable geometry-inspired statistical predictions. 

For each position $\vec{x}$, a DT can be used to obtain the CDF of the received power ${\vec{\psi}}^{\sf{DT}}(\beta, \vec{x}) = P(p(\vec{x}) \leq \beta)$. If the DT is perfectly calibrated then ${\vec{\psi}}^{\sf{DT}}(\beta, \vec{x})$ can be directly used to provide an estimate $q^{\sf{direct}}_{\epsilon}(\vec{x})$. However, in this work, we consider \emph{uncalibrated} DTs generated from open-source maps (e.g. OpenStreetMap \cite{OpenStreetmaps}) and with default material properties. Thus, due to mismatched material properties, changes in the wireless environment and other non-idealities, the direct use of a DT does not necessarily provide an accurate estimate of  $q_{\epsilon}(\vec{x})$. However, the DT's received power CDF ${\vec{\psi}}^{\sf{DT}}(\beta, \vec{x})$ can be leveraged to provide site-specific geometric features.

 We propose to exploit the geometric information implicitly conveyed by ${\vec{\psi}}^{\sf{DT}}(.)$ and a small number of channel measurements to predict fading statistics across the entire site. In particular, by processing the DT's distribution of received power $\vec{\psi}^{\sf{DT}}(\beta,\vec{x})$,  we employ the following predictive distribution
 \begin{align}
 \label{GP proposed}
    q^{\sf{DT}}_{\epsilon} (\vec{x}| \mathcal{D}) &\sim \mathcal{N} (\mu_{\sf{DT}}(\vec{y}| \mathcal{D}), \sigma_{\sf{DT}}^2( \vec{y}| \mathcal{D})), \nonumber
    \\ 
  \vec{y} &= (\vec{x}, \vec{f} (\vec{\psi}^{\sf{DT}}(\beta,\vec{x}))).
 \end{align} 
{Here $\vec{f} (\cdot)$ uniformly samples $100$ points from the DT's CDF of received power $\vec{\psi}^{\sf{DT}}(\beta, \vec{x})$. 

With the $\vec{y}$ defined above, the GP kernel $k(\vec{y}, \vec{y}')$ not only captures spatial patterns but also implicitly encodes geometric information provided by the DT. In particular,  for two receiver positions $\vec{x}_1$ and $\vec{x}_2$ that result in similar DT outputs, $\vec{\psi}^{\sf{DT}}(\beta, \vec{x}_1) \approx \vec{\psi}^{\sf{DT}}(\beta, \vec{x}_2)$,  the GP's covariance function ensures that the predicted channel statistics at these two locations are similar. Thus, the DT's geometric features are captured via the GP's covariance function, embedding geometric correlations that allows efficient prediction of channel statistics using a  limited number of measurements. Moreover, this proposed prediction framework  is shown to provide significant improvements in prediction errors compared to the following \emph{benchmarks}: 
 \begin{itemize}
     \item \emph{benchmark~1:}  direct prediction scheme using $q^{\sf{direct}}_{\epsilon}(\vec{x})$. 
    \item  \emph{benchmark~2:}  spatial prediction scheme $q_{\epsilon}^{\sf{spatial}} (\vec{x}| \mathcal{D})$.  
\end{itemize}

Before presenting the experimental results, which corroborate the efficacy of our framework, the next section  provides theoretical and experimental justification for an effective method for the experimental characterization of small-scale fading statistics.

\section{Experimental Characterization of Small-Scale Fading Statistics}\label{sec:scenario}
{Obtaining spatial samples of the power quantiles  $p_\epsilon\p{\vec{x}}$ requires either relying on model based approaches which are susceptible to model mismatch errors, or a very high number of channel measurements to establish an empirical distribution, which results in high latency. An alternative approach is to use a range of subcarriers and use frequency as a proxy for space \cite{kallehauge2024experimental}. By measuring a wide bandwidth, an empirical channel power distribution can be established and an estimate of the $\epsilon$-quantile of the channel power, $p_\epsilon\p{\vec{x}}$, can be extracted.}

\subsection{Theoretical Analysis}
{The validity of using frequency as a proxy for space can be argued by considering the WSSUS Rayleigh fading channel \cite{parsons_mobile_2000}. In the farfield of the transmitter, the impinging field can be written as a position dependent distribution of planewaves. In quantizing the distribution, the received signal in the vicinity of a point $\veci{x}$ can be written as $h\p{ \vec{x}, f} \approx  \sum_{n= 1}^N \alpha_n \e{-\iu k \vec{x}\T \vec{r}_n} \e{-\iu \tau_n f}$, where $k \in \mathbb{R}$ is the wavenumber, $ \vec{x} \in \mathbb{R}^{3\times 1}$ is a point in the vicinity of $\veci{x}$ and $\alpha_n \in \mathbb{C}$ is the amplitude and phase, $\vec{r}_n \in \left\{ \mathbb{R}^{3\times 1} | \vec{r}_n\T \vec{r}_n = 1 \right\}$ is the normal vector indicating the direction of propagation, and $\tau_n$ is the excess delay of the $n$'th planewave. In the Rayleigh channel, $\alpha_n$ is a set of zero mean independent identically distributed circular symmetric complex random variables. Independence between all $\tau_n$, $\vec{r}_n$, and $\alpha_n$ is assumed. Due to  $\alpha_n$ being circular symmetric complex random variables, $\Aver{h\p{\vec{x}, f}} = 0$. Additionally, due to the central limit theorem,  $h\p{\vec{x}, f} \sim \mathcal{CN}\p{0, \sigma^2}$ for high $N$, where $\sigma^2 = \Aver*{\abs{h\p{\vec{x}, f}}^2}$.
To allow relating the power quantile obtained from sampling across frequency to the power quantile obtained from sampling across space, the covariance properties of the channel across frequency and space, is analyzed. The channel covariance function is given by 
$\Sigma^2\p{\vec{x}_1, \vec{x}_2, f_1, f_2}  = \Aver*{h\p{\vec{x}_1, f_1} h^*\p{\vec{x}_2, f_2}},  
    = \abs{s}^2 \sigma^2    c_\vec{x}\p{\vec{x}_1 - \vec{x}_2} c_f\p{f_1 - f_2},$
where $c_\vec{x}\p{\vec{x}} = \Aver*{\e{-\iu k \vec{x} \T \vec{r}_n}}$ is the spatial correlation function, and $c_f\p{f} = \Aver*{ \e{-\iu \tau_n f}}$ is the frequency correlation function. Assuming a uniform planewave normal distribution, $\vec{r}_n = \begin{bmatrix} \ec{\phi_n} & \es{\phi_n} & 0 \end{bmatrix}\T$ with $\phi_n \sim \mathcal{U}\p{0, 2\pi}$, as in Clarke's two-dimensional model \cite[Sec. 5.4]{parsons_mobile_2000}, and a uniform excess delay distribution $\tau_n \sim \mathcal{U}\p{0, \tau_\text{max}}$, the covariance functions are given as $c_\vec{x}\p{\vec{x}} = J_0\p{2\pi x \lambda^{-1}}$ and $
    c_f\p{f} =  \frac{1 - \e{-\iu \tau_\text{max} f}}{\iu \tau_\text{max} f}$, where $x = \norm{\vec{x}}_2$ is the distance and $\lambda \in \mathbb{R}$ is the wavelength. It can observed that samples are approximately uncorrelated, $c_\vec{x}, c_f < 0.33$, for $x \geq 0.71 \lambda$ and $f \geq 4.6\, \tau_\text{max}^{-1}$. In setting $\tau_\text{max} = 10 d c^{-1}$ with $c$ being the speed of light in a vacuum and $d$ being the distance between transmitter and receiver, the channel is approximately uncorrelated for a frequency shift of $f = 1.4$ MHz at a distance of $100$ m, and approximately uncorrelated for a frequency shift of $f = 6.9$ MHz at a distance of $20$ m. Thus, given sufficient bandwidth and distance, sampling across frequency and space is equivalent of drawing independent random samples from the same distribution.} 
    
\subsection{Numerical Validation}
\label{simulation: statistics of small-scale fading via frequency-domain sampling}
\begin{figure}
\centering
\includegraphics[width= \linewidth, height=4.5cm]{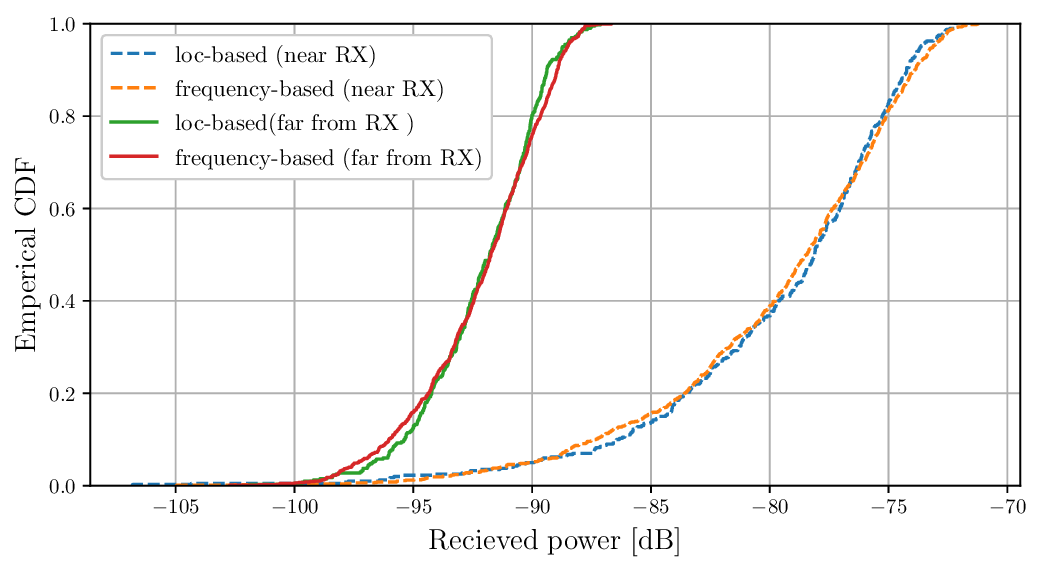}
\caption{{Frequency-based and location-based CDF of received power for two TX positions, one near the receiver (position 55 in Fig.~\ref{fig:scene_layout}) and another far from the receiver (position 111).}}
\label{fig:freq_loc_based}
\end{figure}

For each of the 127 transmitter locations shown in Fig.~\ref{fig:scene_layout}, the small-scale fading distribution is computed by employing the DT to generate a grid of 400 small movements (on the order of a wavelength $\lambda$) in the transmitter's $x$ and $y$ coordinates. Using the channel impulse response (CIR) generated by the DT, an empirical \emph{location-based} distribution of the channel power in dB is computed. The empirical \emph{frequency-based} channel power distribution is also obtained using the DT by fixing the transmitter's position and sampling the frequency domain over a large bandwidth (i.e., 8 GHz, as suggested by \cite{kallehauge2024experimental}).

Fig.~\ref{fig:freq_loc_based} shows the empirical cumulative distribution functions (CDFs) of received power for both the frequency-based and location-based approaches. The results are shown for two typical transmitter positions, one near the receiver and another far from the receiver. As Fig.~\ref{fig:freq_loc_based} shows, the frequency-domain channel power distribution closely approximates the location-based  channel power distribution, including the tails of the distributions. For example, when the receiver is far from the transmitter, the 0.01-quantiles of the location-based and frequency-based distributions differ by only 0.02 dB.  Thus, frequency-domain samples, obtained by wideband channel sounding, offer a practical method to approximate small-scale fading distributions that are otherwise challenging to capture directly.

\section{Experimental results}\label{sec:simulation}
\subsection{Experimental Setup and Frequency Measurements}
\label{sec:experimental_setup}
At each of the 127 transmitter locations shown in Fig.~\ref{fig:scene_layout}, wideband channel measurements were performed in the frequency domain. A total of 8\,001 equally spaced frequency samples were taken per site, spanning the band from $2$GHz to $10$GHz, resulting in a frequency resolution of $1$MHz. During each measurement, the receiver remained stationary while the transmitter was sequentially positioned at the designated locations. Detailed information on the hardware, and measurement protocols can be found in \cite{kallehauge2024experimental}.

\begin{figure}
\begin{minipage}{\linewidth}
\hspace{0.5cm}
\includegraphics[width=0.7\linewidth]{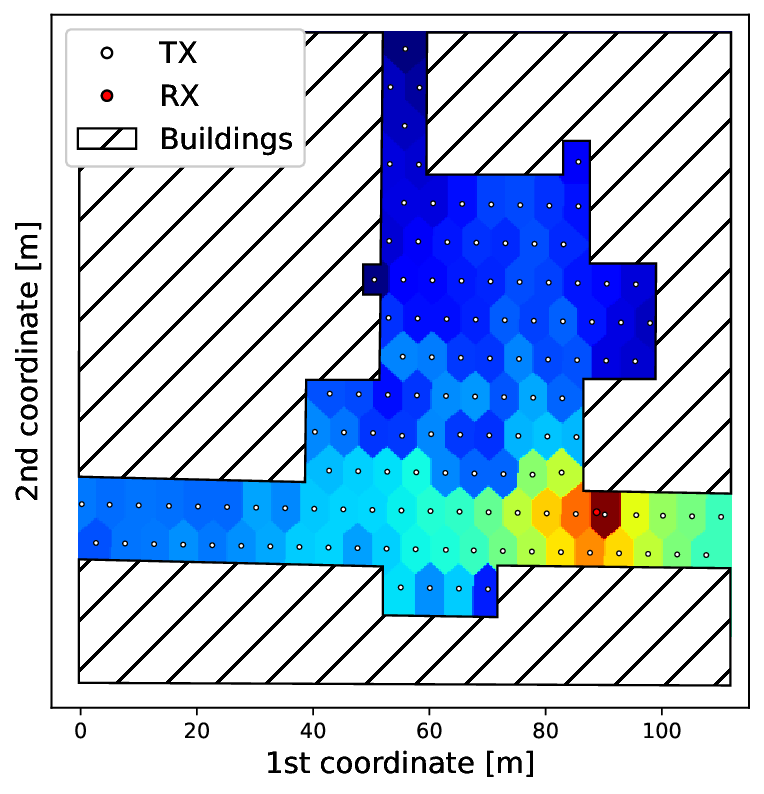}
    \subcaption{Ground truth}
    \label{fig:Ground truth}
\end{minipage}
\begin{minipage}{\linewidth}
\hspace{0.5cm}
\includegraphics[width=0.84\linewidth, ]{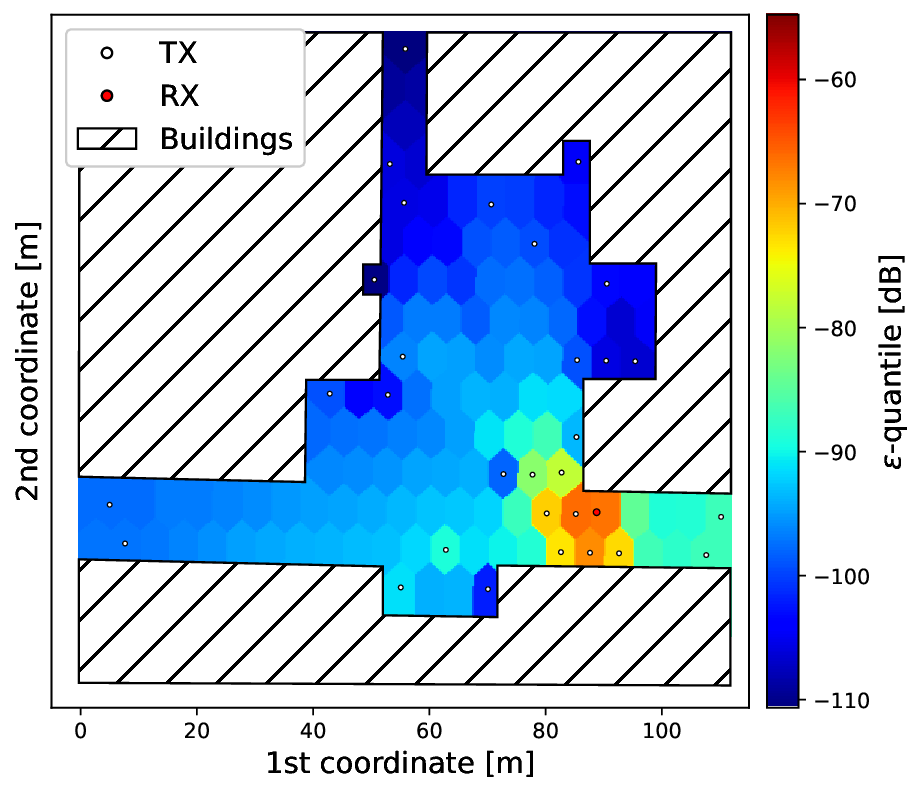}
    \subcaption{Prediction by proposed GP}
    \label{fig:Prediction by (Uncalibrated) DT}
\end{minipage}
\caption{{Closest point interpolation of $\epsilon$-quantile of fading in dB associated with (ground truth) measurements  and predictions by proposed GP and $D=30$ observations.}}
\label{fig:estimated quantiles}
\vspace{-10pt}
\end{figure} 

\subsection{Channel Statistics Predictions}
\label{experiment: statistics prediction}

\begin{figure*}[t]
\centering
\begin{minipage}{0.27\linewidth}
\includegraphics[width=\linewidth]{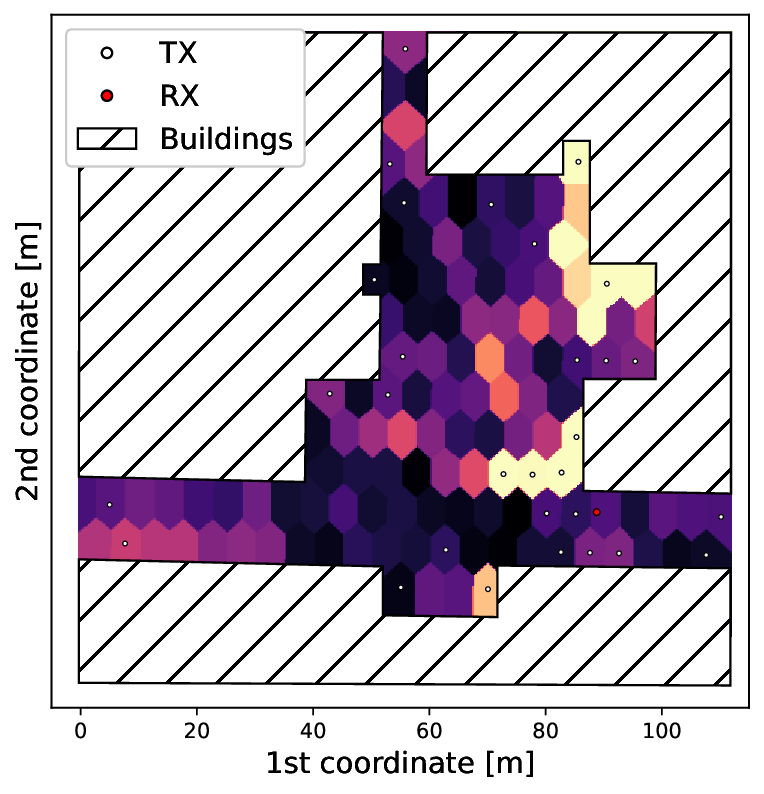}
    \subcaption{}
    \label{fig:Prediction errors RT}
\end{minipage}
\begin{minipage}{0.27\linewidth}
\includegraphics[width=\linewidth]{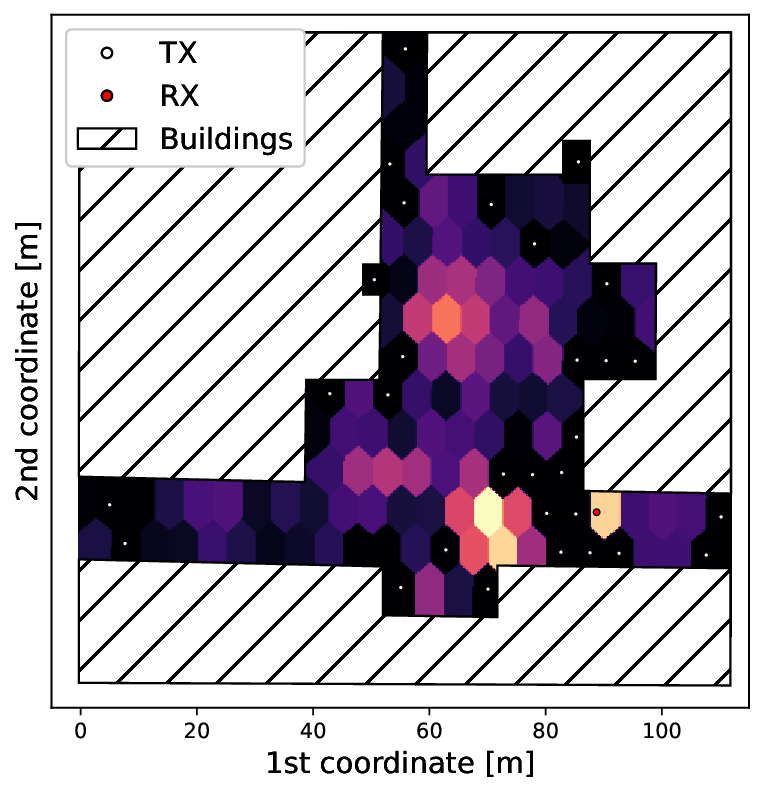}
    \subcaption{}
    \label{fig:Prediction errors baseline GP}
\end{minipage}
\begin{minipage}{0.31\linewidth}
\includegraphics[width=\linewidth]{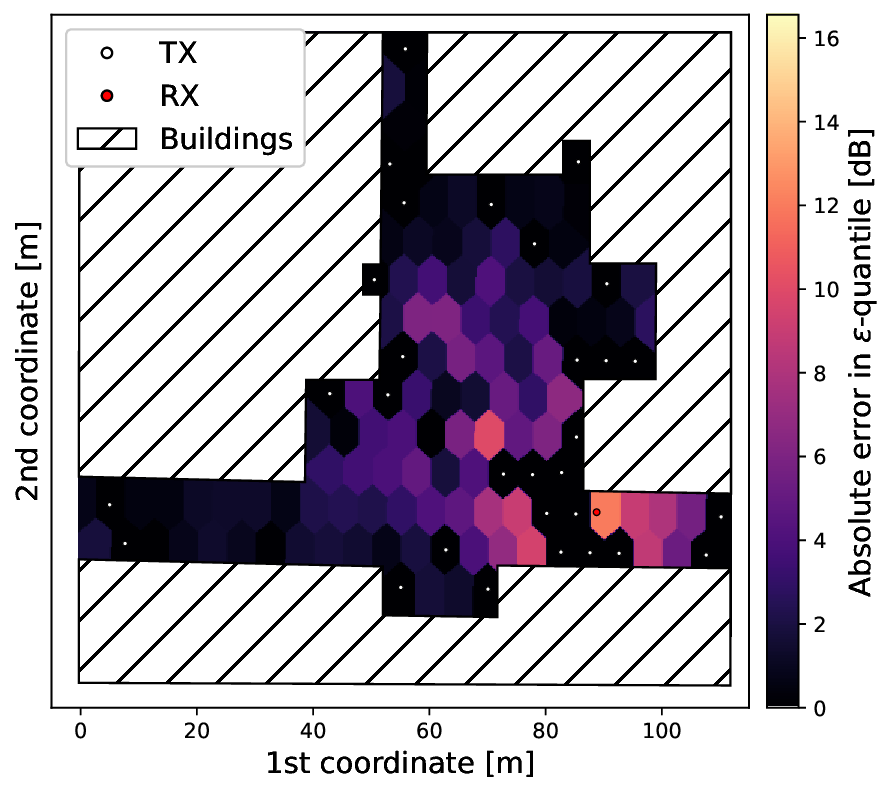}
    \subcaption{}
    \label{fig:prediction errors proposed GP}
\end{minipage}
\caption{{Absolute Error in predicting $\epsilon$-quantile of fading using (a) (uncalibrated) DT (b) baseline GP and (c) Proposed GP.}}
\end{figure*}

\par
In this section, we consider $\epsilon =1 \% $ and the measurement campaign consisting of wideband channel measurements for the 127 positions associated with the setup in Fig.~\ref{fig:scene_layout} which is split into a (small) training set of $30$ positions and an evaluation set of $97$ positions. Here, the  (training) data set is given $\mathcal{{D}}=\{\widehat{q}_{\epsilon,d},\vec{x}_d\}_{d=1}^D$ with $D=30$. The goal is to  predict the $\epsilon$-quantile associated with the remaining $97$ positions.

 We demonstrate the ability of the proposed prediction framework $q^{\sf{DT}}_{\epsilon} (\vec{x}| \mathcal{D}) \sim \mathcal{N} (\mu_{\sf{DT}}(\vec{y}| \mathcal{D}), \sigma^2_{\sf{DT}}( \vec{y}| \mathcal{D}))$ presented in Section~\ref{sec: Site-specific Statistics Prediction}. Fig.~\ref{fig:Ground truth} shows $\epsilon$-quantile of fading under the ground-truth measurements obtained through closest point interpolation. Fig.~\ref{fig:Prediction by (Uncalibrated) DT} shows  $\mu_{\sf{DT}}(\vec{y}| \mathcal{D})$ used to predict $\epsilon$-quantile of fading  with $\mathcal{D}$ containing $30$ observations (corresponding positions are indicated with white dots). To quantify estimation errors, compared to ground-truth measurements in Fig.~\ref{fig:Ground truth},   Fig.~\ref{fig:prediction errors proposed GP} shows the absolute prediction error incurred by  $\mu_{\sf{DT}}(\vec{y}| \mathcal{D})$. For comparison, Fig.~\ref{fig:Prediction errors RT}  and Fig.~\ref{fig:Prediction errors baseline GP} show the   absolute prediction error incurred by the two benchmarks  described in Section \ref{sec: Site-specific Statistics Prediction}: (a) directly using  DT for prediction $q^{\sf{direct}}_{\epsilon}(\vec{x})$ and (b) using spatial features only $q_{\epsilon}^{\sf{spatial}} (\vec{x}| \mathcal{D})$. As shown in the figure, compared to the benchmarks, the proposed method significantly improves prediction accuracy across the entire scene by effectively leveraging the geometric consistency of the DT. For instance, the proposed prediction scheme reduces the median error to $1.8$ dB compared to $3.7$ dB and $3.2$ dB associated with $q^{\sf{direct}}_{\epsilon}(\vec{x})$ and $q_{\epsilon}^{\sf{spatial}} (\vec{x}| \mathcal{D})$, respectively. 

As we will demonstrate next, this improvement in channel statistics prediction can significantly enhance the reliability and performance of a wireless systems.

\subsection{Rate selection under reliability constraints}
\label{section: Rate selection under reliability constraints}

{For a position $\vec{x}$, the predictive distribution $q^{\sf{DT}}_{\epsilon} (\vec{x}| \mathcal{D})$ is employed to select a rate $R^{\sf{DT}} (\vec{x}| \mathcal{D})$ that complies with a meta-probability $\Tilde{p}_{\epsilon}(\vec{x})$ reliability constraint.  In particular,  for a confidence parameter $\delta >0$, the goal is to select  $R^{\sf{DT}} (\vec{x}| \mathcal{D})$ such that the corresponding meta-probability $\Tilde{p}^{\sf{DT}}_{\epsilon}(\vec{x}) \leq \delta$. Following \cite{kallehauge2024experimental}, $R^{\sf{DT}} (\vec{x}| \mathcal{D})$ is selected using the predictive distribution $q^{\sf{DT}}_{\epsilon} (\vec{x}| \mathcal{D})$ given in \eqref{GP proposed}. Here, the rate is selected as the $\delta$-quantile of  $q^{\sf{DT}}_{\epsilon} (\vec{x}| \mathcal{D})$. In particular, for predictive mean $\mu_{\sf{DT}}(\vec{y}| \mathcal{D})$  and predictive variance  $\sigma_{\sf{DT}}^2( \vec{y}| \mathcal{D})$,
\begin{align}
  R^{\sf{DT}} (\vec{x}| \mathcal{D}) = 
  &\log_2 \Big(1 + \e {  \mu_{\sf{DT}}(\vec{y}| \mathcal{D}) + \sqrt{2} \sigma_{\sf{DT}}( \vec{y}| \mathcal{D}) \erf^{-1} (2 \delta-1) } \Big). \nonumber   
\end{align}
The predictive distribution    $q_{\epsilon}^{\sf{spatial}} (\vec{x}| \mathcal{D}) \sim \mathcal{N} (\mu(\vec{x}| \mathcal{D}), \sigma^{2} (\vec{x}| \mathcal{D})$ can be similarly used to select a corresponding rate select  $R^{\sf{spatial}} (\vec{x}| \mathcal{D})$.} To evaluate the performance of $R^{\sf{DT}} (\vec{x}| \mathcal{D})$, we compute its ratio with respect to the $\epsilon$-outage capacity $R_{\epsilon}(\vec{x})$ computed using the (perfect channel knowledge) ground truth fading distribution.  This metric is referred to the \emph{normalized rate} and is given by $\hat{R}^{\sf{DT}}_{\epsilon}(\vec{x}) = \frac{R^{\sf{DT}} (\vec{x}| \mathcal{D})}{R_{\epsilon}(\vec{x})}$.

For $\epsilon=1\%$ and $\delta=10\%$, Fig.~\ref{fig:estimated rates ratio} shows the CDF of $\hat{R}^{\sf{DT}}_{\epsilon}(\vec{x})$ across the $97$ testing positions. For comparison, the figure also shows the CDF associated with the the normalized rate $\hat{R}^{\sf{spatial}}_{\epsilon}(\vec{x}) = R^{\sf{spatial}} (\vec{x}| \mathcal{D}) / R_{\epsilon}(\vec{x})$. As shown in the figure, both $R^{\sf{DT}} (\vec{x}| \mathcal{D})$ and $R^{\sf{spatial}} (\vec{x}| \mathcal{D})$ satisfy the meta-probability reliability criteria. This follows as the proportion of normalized rates exceeding $1$, indicating violations of the outage capacity, is constrained within the $\delta = 10\%$ target. However, the proposed $R^{\sf{DT}} (\vec{x}| \mathcal{D})$ offers a significant boost in the selected rate compared to $R^{\sf{spatial}} (\vec{x}| \mathcal{D})$, as evidenced by the overall increase in normalized rates. Specifically, across different positions, the average normalized rate nearly doubles, improving from $\bar{\hat{R}}^{\sf{spatial}}_{\epsilon}(\vec{x}) = 0.27$ to $\bar{\hat{R}}^{\sf{DT}}_{\epsilon}(\vec{x}) = 0.45$. Consequently, despite limited knowledge of the channel or scene properties, the proposed scheme not only upholds the $\delta = 10\%$ meta-probability guarantee but also significantly reduces the gap to the ideal benchmark that assumes perfect channel knowledge.

\begin{figure}[h]
    \centering
      \includegraphics[width= \linewidth,height=5.5cm]{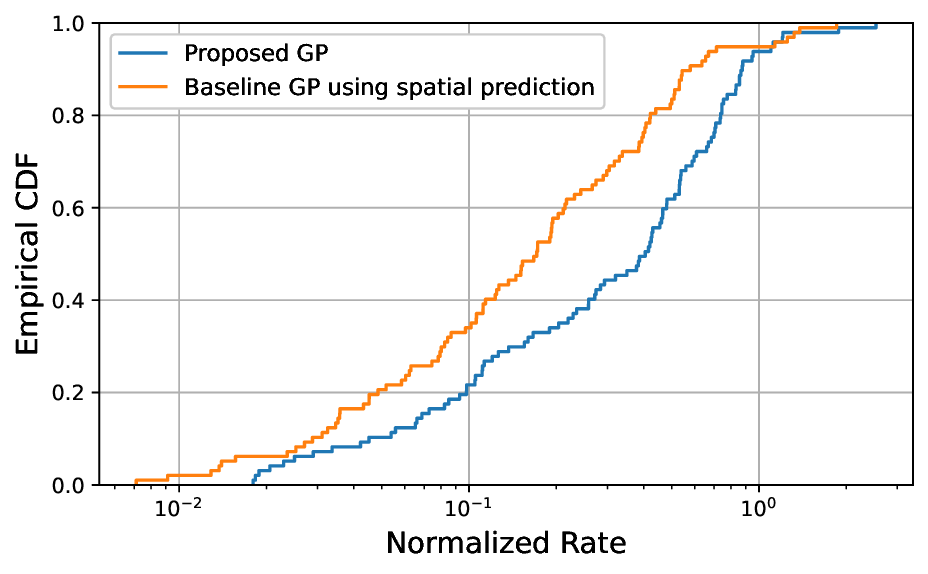}
    \caption{{Empirical distribution of $R^{\sf{DT}} (\vec{x}| \mathcal{D})$ and $R^{\sf{spatial}} (\vec{x}| \mathcal{D})$ normalized by $R_{\epsilon}(\vec{x})$ for  $\epsilon=1 \%$ and $\delta = 10 \%$. }}
    \label{fig:estimated rates ratio}
\end{figure}

\section{Conclusions}\label{sec:conclusion}
In this paper we have shown that by incorporating ray-traced channel information from an inaccurate DT, one can improve the accuracy of channel statistics prediction. The proposed system is capable of almost doubling the provided data rate while maintaining the same reliability, when compared with the baseline. This increase does not come at the expense of increased overhead and lengthy DT calibration procedures, as the system utilizes an uncalibrated DT based on geometry from OpenStreetMap \cite{OpenStreetmaps} with all objects, ground included, having material parameters equal to that of a brick wall. The proposed framework successfully utilizes a DT based on freely available geometry data without calibration to directly improve the performance of a communication system.

\appendices
\bibliographystyle{IEEEtran}
\bibliography{IEEEabrv,references}
\end{document}